\documentclass[useAMS,usenatbib]{mn2e}

\usepackage{epsfig}
\usepackage{lscape} % Allows landscape environment to be used

\input{psfig.sty}

\title[The deprojected 3-D power spectrum of dark and luminous matter]{The three dimensional power spectrum of dark and luminous matter from the VIRMOS-DESCART 
cosmic shear survey}

\author[Ue-Li Pen et al.]{
Ue-Li Pen,$^{1}$\thanks{E-mail:\ pen@cita.utoronto.ca}
Tingting Lu,$^{2}$\thanks{E-mail:\ ttlu@astro.utoronto.ca}
Ludovic van Waerbeke,$^{3}$\thanks{E-mail:\ waerbeke@iap.fr}
Yannick Mellier$^{3,4}$\thanks{E-mail:\ mellier@iap.fr}
\\
$^1$\,Canadian Institute for Theoretical Astrophysics, University of
Toronto, M5S 3H8, Canada \\
$^2$\,Department of Astronomy and Astrophysics, University of Toronto, M5S
3H8, Canada \\
$^3$\,Institut d'Astrophysique de Paris, 98 bis, boulevard Arago, 75014
Paris, France \\
$^4$\,Observatoire de Paris/LERMA, 77 avenue Denfert Rochereau, 75014
Paris, France \\
}

\begin{document}

\date{version 12 April 2003}

\pagerange{\pageref{firstpage}--\pageref{lastpage}} 
\pubyear{2003}

\maketitle

\label{firstpage}

\begin{abstract}

We present the first optimal power spectrum estimation and three
dimensional deprojections for the dark and luminous matter and their
cross correlations.  The results are obtained using a new optimal fast
estimator \citep{2003astro.ph..4513L}, deprojected using minimum
variance and SVD techniques.  We show the resulting 3-D power spectra
for dark matter and galaxies, and their covariance for the
VIRMOS-DESCART weak lensing shear and galaxy
data.  The survey is most
sensitive to nonlinear scales $k_{\rm NL}\sim 1 h$Mpc$^{-1}$.  
On these scales, our 3-D power 
spectrum of dark matter is in good agreement  with the RCS 3-D power 
spectrum found by \citet{2002PhRvD..66j3508T}.
Our galaxy power
is similar to that found by the 2MASS survey, and larger than that of
SDSS, APM and RCS, consistent with the expected difference in galaxy
population.

We find an average bias $b=1.24\pm 0.18$ for the $I$ selected galaxies, and a
cross correlation coefficient $r=0.75\pm 0.23$.  
Together with the power spectra, 
these results optimally encode the entire two point information about
dark matter and galaxies, including galaxy-galaxy lensing.  We address
some of the implications regarding galaxy halos and mass-to-light
ratios.  The best fit ``halo'' parameter $h\equiv r/b=0.57\pm 0.16$,
suggesting that dynamical masses estimated using galaxies
systematically underestimate total mass.

Ongoing
surveys, such as the Canada-France-Hawaii-Telescope-Legacy survey will
significantly improve on the dynamic range, and future photometric
redshift catalogs will allow tomography along the same principles.

\end{abstract}


\newcommand{\be}{\begin{eqnarray}}
\newcommand{\ee}{\end{eqnarray}}
\newcommand{\beq}{\begin{equation}}
\newcommand{\eeq}{\end{equation}}

\section{Introduction}

The recent measurements of the cosmic microwave background
anisotropies have moved physical cosmology into a new era of precision
measurements \citep{2003astro.ph..2209S}.  The cosmic microwave
background perturbations can be cleanly computed from first
principles, and have been measured with high accuracy.  This allows a
clean inference of conditions at the redshift of recombination, $z\sim
1089$.  To complete the picture, a separate measurement of the state
of the universe at lower redshifts is required.  The original WMAP
results used the distribution of optical galaxies as a proxy for the
distribution of total matter.  The dominant uncertainty in such an
exercise is the relationship between galaxies and total mass
\citep{2003astro.ph..2435C}.

A direct measure of the dynamics of total mass is clearly desirable,
as well as a quantitative measure of the relationship between
total mass and visible matter.  Statistical weak gravitational
lensing provides such a handle, and direct measurements
are already providing accuracies on cosmological parameters
such as $\sigma_8$ comparable to indirect galaxy techniques
\citep{2002astro.ph..3134B,2002ApJ...572L.131R,2002ApJ...572...55H,2002A&A...393..369V,2002astro.ph.10604J,2002astro.ph.10213B,2002astro.ph.10450H}.

The gravitational field from the inhomogeneity of the dark matter
distribution disturbs the light from background galaxies and distorts
the apparent images of galaxies. When this distortion is small, it is
called the weak gravitational lensing. Since gravity acts equally on
all particles, gravitational lensing is a complete probe of the total
matter distribution.  Gravity is dominated by dark matter, which does
not involve complicated gas physics, which makes gravitational lensing
relatively straightforward to calculate from analytical models and N-body
simulations \citep{2000ApJ...537....1W}. All these advantages make weak
gravitational lensing a powerful probe of cosmological parameters and
the matter distribution.

So far all analyses of weak lensing data have been parametric by comparing
the observed two dimensional correlations of shear to the predictions
of standard models.  A direct optimal statistical analyses of
the data sets has so far been beyond the scope of computational
resources.  \citet{2002ApJ...567...31P} presented the first angular
power spectra obtained from inversions of the correlations functions.
\citet{2002astro.ph.10213B} did maximum likelihood estimation
on a low signal-to-noise data set.

In this paper, we extract the 3-D dark and luminous matter power
spectrum from the 2-D weak lensing measurements in the VIRMOS-DESCART
survey.  Similar works have been tested successfully in inverting
the 2-D galaxy angular correlations to 3-D galaxy power spectrum
\citep{1990MNRAS.242P..43M,1993MNRAS.265..145B,2000MNRAS.312..774D,2001ApJ...546....2E,2002ApJ...572..140D,2003astro.ph..4005M}.
By measuring the distribution of galaxies and dark matter from the
same survey, we find a direct measure of the cross
correlation \citep{2002ApJ...577..604H}. 

\section{Data}

The VIRMOS-DESCART data consist of four uncorrelated patches (referred
as fields F02, F10, F14 and F22 according to their RA position) of about
4 square-degrees each and separated by more than 40 degrees. The fields
have been observed with the CFH12k panoramic CCD camera, mounted at the
Canada-France-Hawaii Telescope prime focus, over the periods between
January 1999 and November 2001.  The observations and data reduction
have been described in previous VIRMOS-DESCART cosmic shear papers
\citep{2000A&A...358...30V,2001A&A...374..757V,2002A&A...393..369V}.

The observations have been done with the I-band filter available
on the CFH12k camera a with typical exposure time of one hour.
The final cosmic shear catalog contains 392,055 galaxies with
magnitude $I_{AB}>22$ and median $I_{AB}=$23.6.   Several careful
checks have demonstrated that systematic residuals are very
small. However, \citet{2001A&A...374..757V,2002A&A...393..369V} and
\citet{2002ApJ...567...31P} have shown that a B-mode signal still remains
on scales larger than 10 arc-minutes. Its origin is not yet understood.

\begin{figure}
\vskip 3.3 truein
\includegraphics{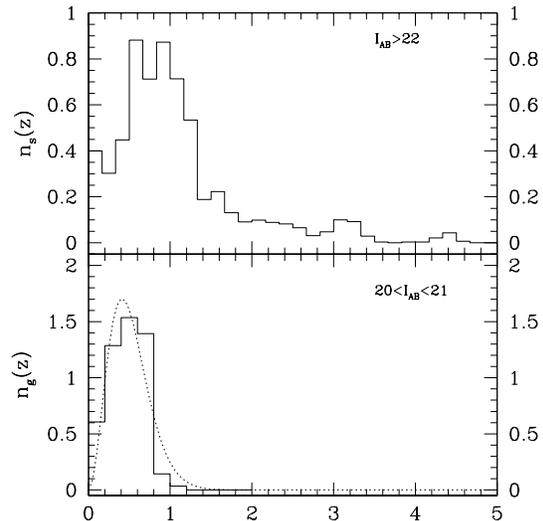}
\caption{The top histogram shows the source weighted redshift distribution of 
the faint $I_{AB}>22$ catalog modeled from the HDF. The bottom histogram
shows the bright $20<I_{AB}<21$ galaxy  
redshift distribution of the catalog modeled from the CFRS (see
section \ref {sec:dep}),  
and the dotted
curve is the fitting formula used in our analyses. 
}
\label{fig:nz}
\end{figure}

The redshift distribution was modeled by the procedure described in
\citet{2001A&A...374..757V} using the photometric redshifts from the
Hubble deep fields (HDF).  Here we performed the same analyses for
all the galaxies with magnitude $I_{AB}>22$. The resulting histogram
for this sample with their appropriate noise
weights is shown in Figure \ref{fig:nz}.
It was modelled from the photometric redshifts of the Hubble
Deep Fields north and South (see \citet{2001A&A...374..757V}).
In the absence of any spectroscopic survey deeper than $I_{AB}=22$
this is the best redshift estimate at the moment.

To compute the galaxy angular power spectra, the galaxy photometries
were calibrated.  A mask file was constructed on a grid spaced
6.18 arc seconds.  For the lens plane sample, we chose a magnitude
range $20<I_{AB}<21$, which contains 20657 galaxies. 

\section{2-D Power Spectrum Estimation}
\label{sec:2d}
Given a reduced galaxy catalog, the first stage in the information
compression process is to produce a two dimensional angular power
spectrum, which encodes all two points information.  Each galaxy has
two observables: a position angle $\theta$ and axis ratio $e$, from
which one forms the two polarization components $e_1=e \cos(2\theta)$
and $e_2=e \sin(2\theta)$.  The data set contains close to a million
degrees of freedom.  A direct optimal power estimation requires ${\cal
O}(N^3)$ operations, which is completely intractable computationally.
One could try to bin the data on a coarse grid to reduce the
computational cost \citep{2000ApJ...537....1W}.  Current processing
power allows one to deploy about 50 grid cells on a side in such a
treatment, corresponding to about 3 arc minute cells.  Such a coarse
binning unfortunately looses information.  Karhunen-Loeve compression
still involves an initially expensive eigenmode expansion, which is
not tractible for a data set of this size.

We applied a novel fast matrix solver, which reduces the problem
from ${\cal O}(N^3)$ to ${\cal O}(N\log N)$ while still being optimal.
\citep{2003astro.ph..4513L}.  The galaxy catalog for each field is binned
onto a square grid 1024 cells wide, and power is estimated directly on
this grid.  The power for the individual fields are then combined by
the procedure of \citet{2002PhRvD..65l3001W}.  The Fisher matrix, error
bars, and window functions are monte-carloed using 1000 realizations.
The angular width of the grid was chosen to be 3.01 degrees, which
is a little larger than the largest dimension of any field.  For the
lowest wave numbers along the longest direction of the grid this
introduces an artificial aliasing due to periodic boundary conditions.
In principle this could be avoided with a two level grid as described
in \citet{2002astro.ph.10478P}, but this has not yet been implemented on
the multi-grid accelerated scheme.  Since we do not use the lowest wave
number bin in our analysis, this should not have a significant effect
on our results.  The analysis is based on a quadratic estimator with
optimal weights relative to a model prior.  We chose the parameters of
the LCDM model in table \ref{tab:models}.  Several strategic choices
are made in the process.  The dark matter and galaxy power spectrum
were estimated independently.  To estimate the cross correlation, both
must be taken into account simultaneously.  When one allows for a prior
correlation of dark matter with galaxies, the eigenmodes of the power are
no longer dark matter and galaxy power, but linear sums and differences.

\begin{table}
\begin{tabular}{rrrrrr}  \hline \hline

Model &$\Omega_m$&$\Omega_{\Lambda}$&$\sigma_8$&$\Gamma$&$h$\\
\hline
LCDM  &0.27 &0.73  &0.9 &0.19 &0.71\\
SCDM1 &1.0  &0.0   &0.5 &0.19 &0.71\\
SCDM2 &1.0  &0.0   &0.5 &0.7  &0.71\\
\hline
\end{tabular}
\caption{Cosmological models used to test the inversion.}
\label{tab:models}
\end{table}

The priors also gave equal E and B mode weights.  The signal is known
to have measurable B mode contamination \citep{2002ApJ...567...31P},
for whose separation a symmetric weight seemed most robust.  Unequal
priors lead to different window functions for the two modes, making
comparisons tricky.  A further potential complication is the additive
white noise calibration in the power spectrum.  As described in
\citet{2002ApJ...567...31P}, the intrinsic distribution of the
unlensed ellipticity distribution can not be measured, so the two
point shear correlation function at zero lag is completely unknown.
This translates into an additive white noise factor in the power
spectrum.  Normally the noise is subtracted from the power spectrum
estimator, and in our case, we estimate the noise by randomly rotating
galaxies.  This will cancel the correlation function at zero lag
(which is invariant under rotations), and determine a fixed value of
the integration constant.  At large $l$ this can lead to a systematic
underestimate of the power.

The reduced catalogues also have a statistical weight for each galaxy,
which in Gaussian analyses is equal to the inverse noise variance.  A
random rotation would assign a noise equal to the observed
ellipticity, but such a procedure biases the estimated power spectrum.
Instead, the noise was assigned using the table lookup described by
\citet{2000A&A...358...30V}.  Our power spectrum analysis thus has a
different procedure for the noise and weights.

\begin{figure}
\vskip 3.3 truein
\includegraphics{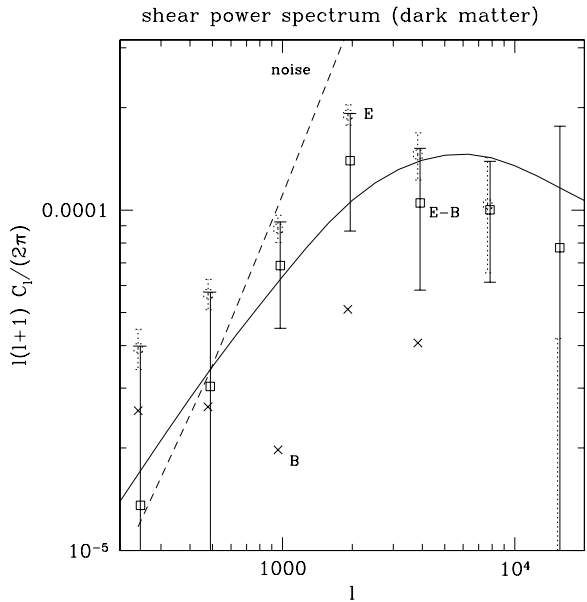}
\caption{Spherical harmonic power spectrum.
The dashed boxes indicate the raw E-type power-spectrum, while 
the crosses denote the B-type power. The solid boxes with error bars are
$E-B$, which is the quantity we used to calculate the three dimensional
power. Their error bars are the quadrature sum
$\sqrt{{\sigma_E}^2+B^2}$.
We have shifted the solid boxes with error bars to larger l slightly, in order 
to be distinguished with E-type and B-type power. 
The dashed straight line is the noise. The solid curved line is the 
model prediction for LCDM (see table \ref{tab:models}).
}
\label{fig:clvirmos}
\end{figure}

Figure \ref{fig:clvirmos} is the power spectrum of cosmic shear, which 
represents the fluctuations in the projected surface density
\begin{equation}
\kappa(\hat{\bf n})=\frac{\Sigma(\hat{\bf n})}
{\Sigma_{cr}},
\end{equation}
where
\begin{equation}
\Sigma_{cr}=\frac{c^2}{4\pi G}\frac{D_s}{D_d D_{ds}}.
\end{equation}
$\Sigma(\hat{\bf n})$ is the surface density, $\hat{\bf n}$ is the
direction on the sky.
$D_d$, $D_{ds}$ and $D_s$ are the angular-diameter distance between
the observer and lens, lens and source, observer and source respectively.

The dashed boxes indicate the E-type power-spectrum, while the crosses
denote the B-type power. The solid boxes with error bars are the
subtracted power $E-B$, which is the quantity we used to calculate the
three dimensional power. For the difference powers, the error bars are
calculated by the quadrature sum $\sqrt{{\sigma_E}^2+B^2}$.  The $B$
mode is taken as a diagnostic for the error estimate, and we only
added the value to the diagonal of the covariance matrix.
Correlations between scales are not accounted for.  The dashed
straight line is the noise, which dominates at small angular scales.
When noise dominates, the inverse noise weighted two point correlation
function is an optimal estimator.  The solid curve line is the power
spectrum projected by the Limber equation from the
three dimensional power using the code by \citet{2002astro.ph..7664S}.
We note that the errors are dominated by the $B$ mode, and eyeballing
the plots, one could see up to three independent useful power spectrum
measurements. 

\begin{figure}
\vskip 3.3 truein
\includegraphics{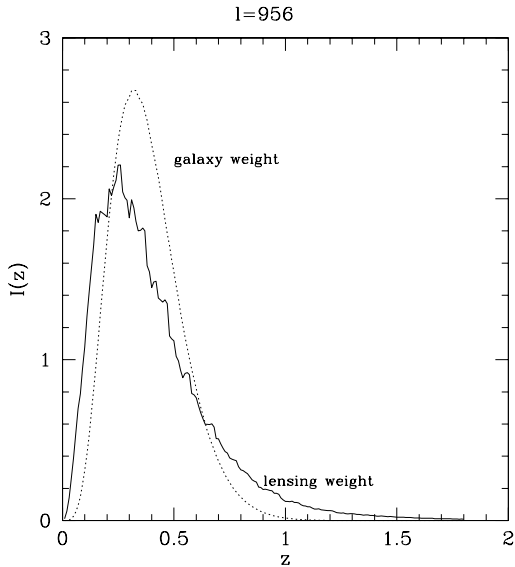}
\caption{
The contribution to 2-D power spectrum by the 3-D power spectrum from different 
redshifts for a fixed angular scale $l=956$. The dotted line is for
the galaxies. The solid line is for 
the  dark matter.
}
\label{fig:peak}
\end{figure}

\begin{figure}
\vskip 3.3 truein
\includegraphics{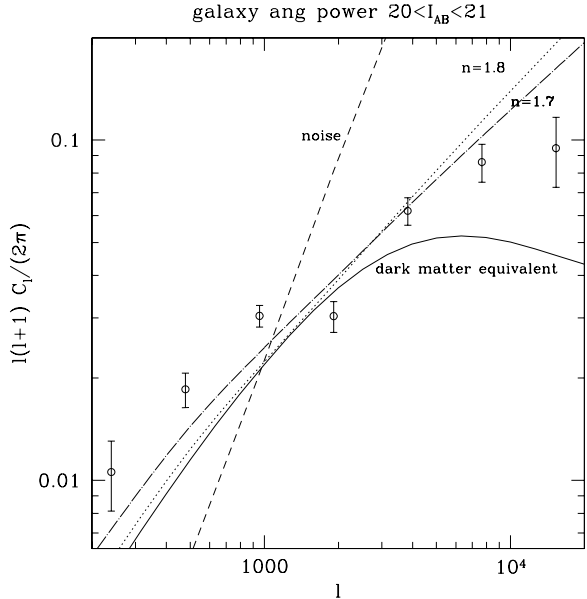}
\caption{The circles indicate the power spectrum for galaxies $20<I_{AB}<21$.
The dashed line is the Poisson noise, and the solid line is the
Limber equation prediction for  luminous
matter perfectly tracing dark matter.
The dot-dashed and dotted lines are the projection of the best fit
power spectra parametrized as
$\Delta^2=(k/k_0)^n$, for $n=1.7$ and 
$k_0 = 0.16 (h \rm Mpc^{-1})$, as well as
$n=1.8$ and $k_0 = 0.18 (h \rm Mpc^{-1})$ respectively.
}
\label{fig:clvgal}
\end{figure}

To measure the distribution of luminous matter, we chose a magnitude
range which traces the redshift distribution of lenses. The range
$20<I_{AB}<21$ results in a differential redshift contribution shown in
Figure \ref{fig:peak}, which is well matched to the lens weights.  We
used this magnitude range to measure the galaxy power spectrum
shown in Figure \ref{fig:clvgal}.

In the two dimensional projection, we fitted for a parametrized power
spectrum $\Delta^2=(k/k_0)^n$.  This corresponds to a power law
correlation function of the form $\xi=(r/r_0)^{-n}$, with $r_0=2
\left[\pi/2\sin(n\pi/2) \Gamma(2-n)\right]^{1/n}/k_0$.
We used two values of $n$: 1.8 and 1.7.  For $n=1.8$, the best fit
correlation length is $k_0=0.22\pm 0.010 h$Mpc$^{-1}$ for the $I$
selected galaxy population at weighted redshift $z_m=0.36$.  The
corresponding correlation length is $r_0=4.79\pm 0.22 h^{-1}$ Mpc.  
The shallower slope $n=1.7$ fits
the data slightly better, and results in a 17\% longer correlation
length.  The two dimensional galaxy power certainly appears consistent
with expectations for this population of galaxies.

\begin{figure}
\vskip 3.3 truein
\includegraphics{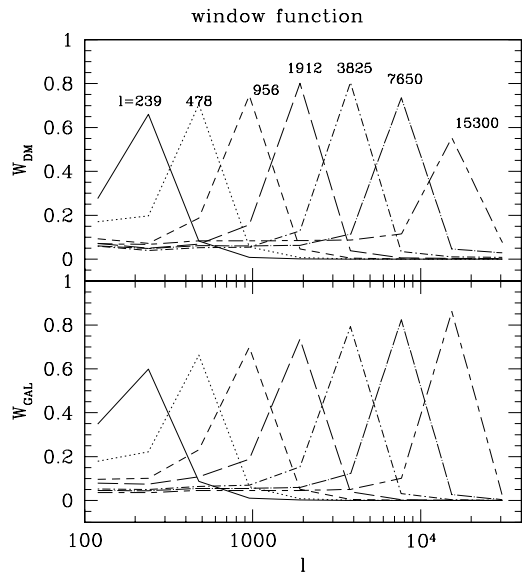}
\caption{
The window function for the two dimensional power spectrum of shear
and galaxies.  
>From left to right, the lines are for
$l$=239, 478, 956, 1912, 3825, 7650, 15300 respectively.
The windowed variable is $l(l+1) C_l/2\pi$, so for flat
band-power more power leaks from large scales to small than vice
versa.}
\label{fig:clwindow}
\end{figure}

We used prior weights in the galaxy power spectrum estimation
corresponding to the $n=1.7$ dot-dashed line in Figure
\ref{fig:clvgal}. 
The galaxy power spectrum estimation is accomplished on a significantly masked
geometry, which results in aliasing of modes.  This is described by
the window function, shown in Figure \ref{fig:clwindow}.

The finite sampling of the 1000 Monte-Carlo simulations to compute the
Fisher matrix results in an expected error of about 4\% on the diagonals.  
On the off-diagonals, this error becomes large compared to the actual values
after about the second distant bin, since the actual correlations become
quite small.  We only included the diagonal and the first off-diagonal
of the Fisher matrix in all subsequent analyses.

Computing the cross correlation we initially used a light-traces-mass
prior.  The quadratic estimator just weights the shear and galaxy
surface density individually by a Wiener filter.  For a perfectly
correlated galaxy and dark matter field, the expected cross
correlation coefficient is 0.96, indicating that the lensing weights
and galaxy weights overlap very strongly.

\begin{figure}
\vskip 3.3 truein
\includegraphics{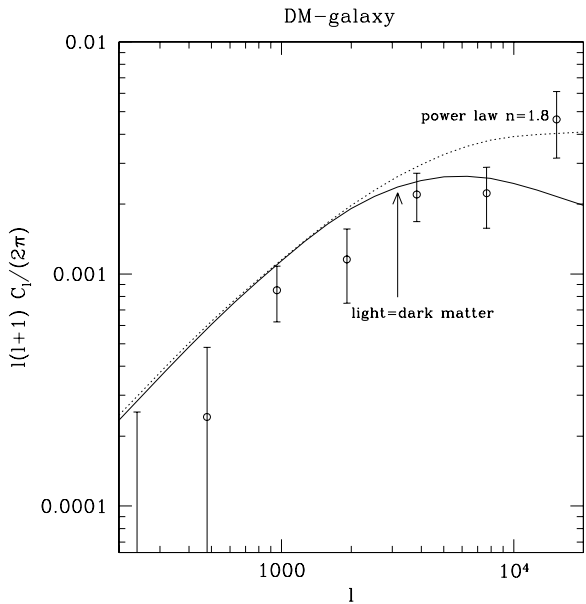}
\caption{The cross correlation between dark matter and galaxies, also
called galaxy-galaxy lensing power.
The circles indicate the power spectrum for galaxies $20<I_{AB}<21$.
The solid line is the Limber projection for galaxies tracing dark
matter perfectly.
The dotted line is the projection of a power law galaxy correlation
(see text) with $n=1.8$, which has no stochasticity relative to dark matter.
}
\label{fig:cldg}
\end{figure}

The resulting power spectrum is shown in Figure \ref{fig:cldg}.  It is
apparent that the cross correlation is systematically lower than
expected for a perfectly non-stochastic galaxy distribution (dotted line).

\section{Deprojection}
\label{sec:dep}

The projection of a three dimensional power spectrum 
\begin{equation}
\Delta^2(k,z)\equiv\frac{k^3}{2\pi^2}P(k,z)
\end{equation}
to a two dimensional angular power spectrum $l(l+1)C_l/2\pi$
is given by 
Limber's equation \citep{2002PhRvD..65f3001H},
\begin{equation}
\frac{l(l+1)}{2\pi} C_l=\frac{\pi}{l}\int_0^{z_s} \Delta^2(l/\chi(z),z)
w(z)^2 \chi(z)
\frac{d\chi}{dz} dz.
\label{eqn:limber}
\end{equation}
The comoving angular diameter distance is
\begin{equation}
\chi(z)=c\int_0^z {\frac{dz}{H(z)}}
\end{equation}
where H(z) is the Hubble constant at redshift z:
\begin{equation}
H(z)=H_0[(1+z)^2(\Omega_m z+1)-\Omega_{\Lambda}z(z+2)]^{1/2}.
\end{equation}
For the angular diameter distance $\chi$ we used the fitting formula from
\citet{1999ApJS..120...49P}.
The weight function $w(z)$ is defined below for the different
categories of dark and luminous matter.  For broad weight
functions, Limber's equation (\ref{eqn:limber}) 
mixes modes significantly.  A direct deprojection is
analogous to a deconvolution, which is in general numerically
unstable.  Several alternative exist to recover the underlying three
dimensional power spectrum.

All measurements of power spectra, even with full three dimensional
information, measure the power spectrum convolved with a window
function.  One can think of the two dimensional measurement an extreme
case of such a window function.  Each angular wave number $l$ is a sum
over power spectra at different linear wave numbers $k$.  This sum can
be treated analogous to a window function.  The current fashion is not
to attempt to deconvolve power spectra.

\citet{1998ApJ...506...64S} proposed such a minimum variance
procedure, which estimates power in broad bands, but with small
errors.  This approach leads to estimates of power over broad windows,
and is quite robust.  There is a slight model dependency in the
normalization of the window functions, for which we calibrate relative
to a fiducial model described below.  As long as the power spectrum
has a similar shape to the underlying model, the total amplitude is
unbiased.  If the bins are chosen narrower than the windows, they just
become more and more correlated, but error bars remain constant.  We
call the minimum variance procedure Method A.  If one considers Limber's
equation (\ref{eqn:limber}) to be a linear mapping $L$, the minimum
variance procedure weights each data point by the inverse noise
variance, and projects it back using the transpose of $L$.  This gives
a unique parameter-free lossless mapping, and a natural way of
deprojecting the angular wave numbers $l$ back to linear wave numbers
$k$.

This minimum variance deprojection is an even more convolved version
of the original power spectrum.  One can now attempt to deconvolve
this deprojection to narrow down the window function.  As in the past
literature, we stabilize this deconvolution using a singular value
decomposition.  An unbounded inversion has window functions that are
delta functions, and does not need a model to normalize.  In practice,
one has to introduce a cutoff, which reduces power and will always
bias the answer low.  In this case, one can recalibrate relative to a
model.  The number of modes that are chosen is another new free
parameter.  Since significant effort had been spent on this procedure
historically, we include such an analysis for comparison.

We now proceed to describe the specific deprojection for the
VIRMOS-DESCART angular power spectra.  We modelled the non-linear
power spectrum $\Delta^2$ using the \citet{2002astro.ph..7664S} code.
$\Omega_m$ and $\Omega_{\Lambda}$ denote the matter density and
cosmological constant density today.  Whenever a quantity is redshift
dependent, we explicitly include that, for example $\Omega_m(z)$ and
$\Omega_{\Lambda}(z)$ are the values at redshift z.

For dark matter the lensing weight is
\begin{equation}
w_{dm}(z)=\frac{3}{2}\Omega_m {H_0}^2g(z)(1+z)
\end{equation}
where
\begin{equation}
g(z)=\chi(z)\int_z^{+\infty}dz'n_s(z')\frac{\chi(z')-\chi(z)}{\chi(z')}.
\end{equation}
$n_s(z)$ is the distribution of source galaxies, for which we use the
statistically sampled model from the HDF shown in Fig. \ref{fig:nz}:

For galaxies,
\begin{equation}
w_{galaxy}(z)=H(z)n_g(z) 
\end{equation}
where $n_g(z)$ is the lens plane galaxies distribution. 
We used a galaxy distribution parametrized as

\begin{equation}
n_g(z)=\frac{\beta}{z_s \Gamma \left( \frac{1+\alpha}{\beta} \right)}\left(\frac{z}{z_s}\right) ^{\alpha}\exp{\left[-\left( \frac{z}{z_s}\right)^
{\beta}\right]}.
\label{eqn:ngz}
\end{equation}
with fixed values $\alpha=2$ and $\beta=1.5$.
$z_s=0.340$. We modelled $z_s=0.340$ from the CFRS
catalogue \citep{1995ApJ...455...50L,1995ApJ...455...60L,1995ApJ...455...75L,1995ApJ...455...88H,1995ApJ...455...96C}. 
We used the redshift distribution of galaxies with
$20<I_{AB}<21$, discarding objects of class greater than 9 or less
than 2, for which redshift determinations could have been problematic.
This left us with 140 galaxies, with median redshift
$z_M=0.480$.
The uncertainty of $z_M$ is deduced by bootstrapping the catalog
$1000$ times,  resulting in a bootstrap error $\Delta z_M=0.021$. The
bootstrap error may well underestimate the true error, so we applied
the same estimators to each of the five fields separately.
The resulting median redshifts and bootstrap errors are:
0000-00 field\citep{1995ApJ...455...60L}:
$0.195\pm 0.083$,
0300-00 field\citep{1995ApJ...455...88H}:
$0.543\pm 0.039$,
1000+25 field\citep{1995ApJ...455...60L}:
$0.495\pm 0.031$,
1415+52 field \citep{1995ApJ...455...75L}:
$0.519\pm 0.094$, 
and 
the 2215+00 field (ibid): 
$0.341\pm 0.028$
respectly. Combining the results of those patches weighted
by their inverse variances, we find a mean median redshift $0.431\pm
0.018$, which is marginally consistent with the bootstrap error.
Taking the difference squared between the sample median and each field median,
dividing by the sum of the bootstrap variances, we find a $\chi^2=5.8$
per degree of freedom, indicating that the bootstrap errors are
probably an underestimate, and that the
true error is 2.4 times larger if taken at face value.  Unfortunately,
the standard deviation of this variance estimator from 5 fields is
$\sqrt{2/(n-1)}$ = 71\%, so the true error is very poorly known.

In terms of the parameterization (\ref{eqn:ngz}), we have
$z_s=0.340$.     
For the cross correlation between galaxies and dark matter, we changed the
$w(z)^2$ in equation (\ref{eqn:limber}) to the product of dark matter
weight function and galaxy weight function $w_{dm}(z)w_{galaxy}(z)$.

We assumed that the three dimensional power spectrum evolves linearly with 
redshift relative to a reference redshift $z_m$, 
\begin{equation}
\Delta^2(k,z)=\Delta^2(k,z_m)\left(\frac{D(z)}{D(z_m)}\right)^2.
\end{equation}
The errors introduced by this simplification are quantified below.
We chose  $z_m$ to be the redshift below which half the power
originates at the middle 
of our angular scales of interest, $l\sim 1000$.  Writing the contributions
in terms of an integrand $I(z)$,
\begin{equation}
\frac{l(l+1)C_l}{2\pi}=\int_0^{z_s}I(z)dz.
\end{equation}
we find from Figure \ref{fig:peak} that a value $z_m=0.36$ 
is close to the median contribution for both the dark matter and
galaxy distribution.

$D(z)$ is the linear growth factor, for which we use the fitting formula
\citep{1992ARA&A..30..499C}: 
\begin{equation}
D(z)=\frac{1}{1+z}\frac{D_1(z)}{D_1(0)},
\end{equation}
\begin{eqnarray}
D_1(z)&=&\frac{5\Omega_m(z)}{2}[(\Omega_m(z))^{4/7}-\Omega_{\Lambda}(z)\nonumber\\
&& +(1+\Omega_m(z)/2)(1+\Omega_{\Lambda}(z)/70)]^{-1}.
\end{eqnarray}

We group the two-dimensional power spectrum $l(l+1)C_l/2\pi$  into an $n_l$
dimensional vector $\bf y$ and the three-dimensional power spectrum  
$\Delta^2(k,z)$ an $n_k$ dimensional 
vector $\bf x$.  Discretizing the integral (\ref{eqn:limber}) into a
trapezoidal rule sum of 15 redshift slices with $\Delta z=0.12$,
Limber's equation can be written as
\citep{2002PhRvD..65l3001W}
\begin{equation}
\bf y={\bf C}{\bf x}+{\bf n}.
\label{eqn:discretelimber}
\end{equation}
In our analysis, the projection matrix ${\bf C}$ also contains the
effects of the window arising from the power spectrum estimation on
the irregular grid.
To relate the linear analysis of power spectra to that of Gaussian
random fields, we introduced a random noise vector $n$, whose
covariance is defined to be the Fisher matrix:
\begin{equation}
\langle {\bf y y}^t\rangle-\langle {\bf y}\rangle\langle {\bf y}^t\rangle\equiv
{\bf F}_{ij}\equiv\langle {\bf n}{\bf n}^t\rangle.
\end{equation}
We used a bilinear interpolation in $\log k$  to evaluate the power
spectrum in the integrand.

An linear estimate of $x$ can be written as:
\begin{equation}
\tilde{\bf x}={\bf P}{\bf y}.
\end{equation}
When all equations are invertible, one could formally use
\begin{equation}
{\bf P}={\bf C}^{-1}.
\label{eqn:pinv}
\end{equation}
In general, however, $n_l \ne n_k$, and the matrix ${\bf C}$ is not
square, not invertible, and even if it were square, is very ill conditioned.
Then the choice of $\bf P$ becomes important. We will discuss three
procedures
\citep{1997ApJ...480L..87T,1997PhRvD..55.5895T,2001PhRvD..64f3001T,2002PhRvD..65l3001W}:
Method A, which gives minimal but correlated error bars 
\citep{1998ApJ...503..492S,1998ApJ...506...64S} is:
\begin{equation}
{\bf P}_1={\bf N}_1{\bf C}^t{\bf F}^{-1}.
\label{eqn:m1}
\end{equation}
A diagonal normalization matrix ${\bf N}_1$ is defined relative to a
fiducial input model power spectrum ${\bf y}$ such that ${\bf  y}={\bf
  P}_1{\bf y}$. 
In general, one can measure power spectra and cross power spectra as a
function of source redshift, so for each $l$ there could be multiple
measurements of $C_l$ with appropriate covariances ${\bf F}$.  Due to
the limited signal-to-noise and absence of detailed source redshift
information in our survey, we only used one combined power spectrum.
The procedure is general, and can combine any number of source and
cross powers optimally.

Method C is mathematically equivalent to equation (\ref{eqn:pinv}): 
\begin{equation}
{\bf P}_3={\bf N}_3({\bf C}^t{\bf F}^{-1}{\bf C})^{-1}{\bf C}^t{\bf F}^{-1}
\end{equation}
if ${\bf C}^t{\bf F}^{-1}{\bf C}$ is invertible. 
The normalization
${\bf N}_3$ is defined as for (\ref{eqn:m1}).  If the central term is
invertible, the normalization ${\bf N}_3$ is the identity matrix.
We will use SVD 
to stabilize the problem for the general case.  

The basis of SVD comes from the following linear algebra result:
Any $m\times n$ matrix $\bf F$ ($m\geq n$)  can be decomposed to an $m\times n$
column-orthogonal matrix $\bf U$, an $n\times n$  diagonal matrix $\bf W$ with
positive or zero elements(the singular values), and the transpose of an
$n\times n$ orthogonal matrix $\bf V$.
\begin{equation}
{\bf F}={\bf U}{\bf W}{\bf V}^t.
\label{eqn:svd1}
\end{equation}
The matrices $\bf U$ and $\bf V$ satisfy:
\begin{equation}
{\bf U}^t{\bf U}={\bf V}^t{\bf V}^t={\bf 1}.
\label{eqn:svd2}
\end{equation}

If the matrix $\bf F$ is square,then $\bf U$, $\bf V$, and $\bf W$ are all
$n\times n$
square matrices. Then the inverse of $\bf F$ is
\begin{equation}
{\bf F^{-1}}={\bf V}[{\rm diag}(1/w_j)]{\bf U}^t.
\label{eqn:svd3}
\end{equation}
But if one of the $w_j$ is zero, or (numerically) so small that its value is
dominated by roundoff error, the inverse process will be 
incorrect.  SVD prescribes the inverse of these ``singular values'' to
be set to zero.   We use the routines from Numerical Recipes \citep{1992nrfa.book.....P}
to implement this SVD decomposition.

In order to solve our problem:
\begin{equation}
{\bf C}{\bf x}_0={\bf y},
\label{eqn:svd4}
\end{equation}
we first consider the following sets of linear equations:
\begin{equation}
{\bf F}{\bf x}={\bf b},
\label{eqn:svd5}
\end{equation}
and try to invert such a equation for a square $n\times n$ matrix $\bf F$, 
and vectors $\bf x$ and $\bf b$.
The first question is whether $\bf b$ lies in the range of $\bf F$ or not. 
If it does, then the set of equations does have one more more
solutions which may be degenerate.
If $\bf b$ does not lie in the range, then there is no solution.
In both cases, replace those $1/w_j$ by zero if $w_j=0$ or
$w_j$ is very small.  We quantify smallness below.   We then calculate
\begin{equation}
{\bf x}={\bf V} [{\rm diag}(1/w_j)]{\bf U}^t b.
\label{eqn:svd6}
\end{equation}
Here $\bf V$, $\bf W$, and ${\bf u}^t$ are decomposed by $\bf F$. This is the 
solution given by SVD.

In the case of fewer equations than unknowns($m<n$), the method is
also applicable, however there will be an $n-m$ dimensional family of
solutions.  We have to choose a parameter for the threshold to zero
those small $w_j$. Different cutoffs may lead to different solutions.
To find a suitable criteria for the cutoff, we can calibrate with
simulation data.

Finally, an intermediate choice between methods A and C is method B:
\begin{equation}
{\bf P}_2={\bf N}_2 ({\bf C}^t{\bf F}^{-1}{\bf C})^{-\frac{1}{2}}{\bf
  C}^t{\bf F}^{-1}. 
\end{equation}
One might expect that an SVD cutoff is not needed for this case, since
any large 
eigenvalues on the square root are canceled by small eigenvalues of
the last term.

\section{Results}
\subsection{dark matter power spectrum}

The deprojected three-dimensional 
power spectrum $\tilde{\bf x}$ from the observed
angular power  
is given by a linear relation
$\tilde{\bf x}={{\bf P}}{\bf y_{ob}}$, with covariance
\begin{equation}
\langle {\bf \tilde{x} \tilde{x}}^t\rangle-\langle {\bf
  \tilde{x}}\rangle
\langle {\bf
  \tilde{x}}^t\rangle = {{\bf P}}^t{\bf F}{{\bf P}}.  
\end{equation}
$\bf y_{ob}$ denotes the observed
data, written as a $n_l$ dimensional vector.  We used $n_l=7$
starting at l=236 with logarithmic bins each a factor of two wide.
In the three dimensional space we chose wave numbers corresponding to
$k=l/\chi(z_m)$, and added two more bins on each side for a total of
$n_k=11$. 

\begin{figure}
\vskip 3.3 truein
\includegraphics{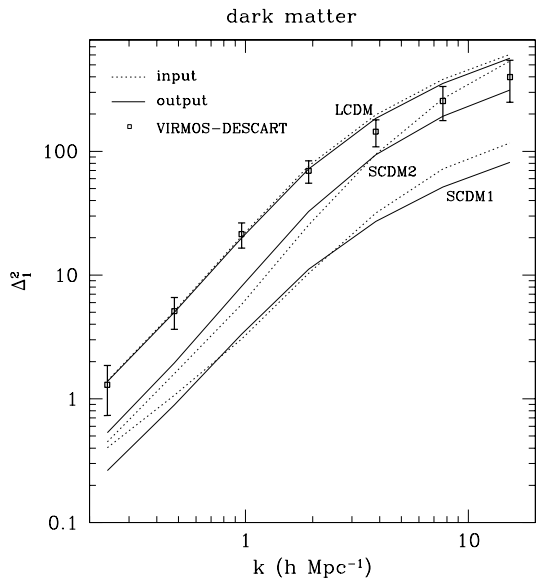}
%\plotone{dpwsa.eps}
\caption{
The 3-D power spectrum of dark matter linearly evolved to redshift
zero, for the minimum variance method A.
The dotted lines are the input power spectrum, linearly evolved from
the \citet{2002astro.ph..7664S} power spectrum at $z_m$. From bottom to
top are cosmologies SCDM1, SCDM2 and LCDM (see table \ref{tab:models}).  
The solid
lines are the deprojected 3-D power spectrum from the non-linearly projected
2D power spectrum.  The boxes are the power spectrum deprojected from
the measured angular power spectrum.
The deprojected lines of SCDM here were scaled with by
$(\Omega_{\rm SCDM}/ \Omega_{\rm LCDM})^{1.2}$ to compensate the
linear evolution difference. 
}
\label{fig:dpwsa}
\end{figure}

The minimum variance solution for Method A is shown in
Fig.\ref{fig:dpwsa}. Only the seven
wave numbers in the central range are plotted.
Error bars are the square roots of the diagonal
elements of the matrix 
\begin{equation}
{\bf E}={{\bf P}}^t{\bf F}{{\bf P}}.  
\label{eqn:covar}
\end{equation}

\begin{figure}
\vskip 3.3 truein
\includegraphics{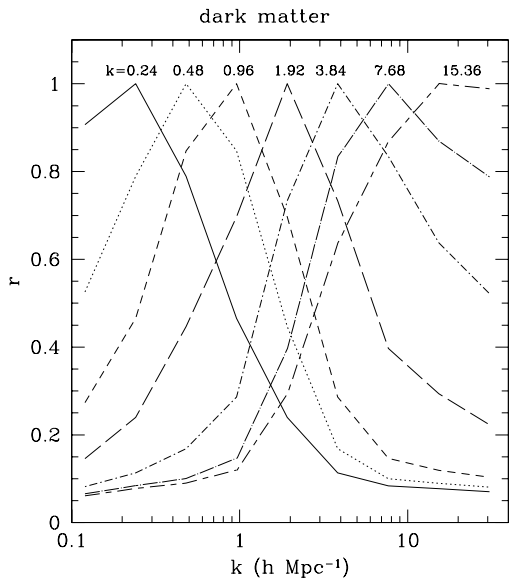}
%\plotone{dcova.eps}
\caption{The cross correlation coefficient betweens bins of Figure
  \ref{fig:dpwsa}.
>From left to right, the covariances are relative to the bin at  
$k=0.24, 0.48, 0.96, 1.92, 3.84, 7.68, 15.36 (h \rm Mpc^{-1})$ respectively.
}
\label{fig:dcova}
\end{figure}

The off-diagonal terms of the covariance matrix can be plotted in terms
of cross-correlation coefficients, 
\beq
r_{ij}\equiv \frac{E_{ij}}{\sqrt{E_{ii}E_{jj}}}.
\eeq
In Figure \ref{fig:dcova} we show the cross correlation coefficient for
the seven central solutions of Figure \ref{fig:dpwsa}.  The points are
clearly significantly correlated.

To simplify the inversion process we assumed linear evolution of the
power amplitude with redshift. We only
specified the 3-D power spectrum at one independent redshift $z_m$.
The inversions are all normalized relative to the prior LCDM cosmology
with linear evolution, for which by construction the resulting 3-D
power spectrum will be the input power.  If we apply the linear
process to the non-linearly projected angular power spectrum, the
results could differ. Using the \citet{2002astro.ph..7664S} code we
can also generate the full non-linear power spectrum at each redshift,
from which we project the two dimensional angular power.  In Figure
\ref{fig:dpwsa} we plotted the input 
power spectrum at $z_m$ as the topmost dotted line, and the linearly recovered
power spectrum as the corresponding solid line.  We find good
agreement, certainly 
better than the other sources of statistical error.

We need to know the stability to a change of cosmological parameter
priors. To check cosmologies that are quite different from
the previously assumed LCDM model, we have also checked the SCDM cosmologies 
listed in table 1. In
Fig.\ref{fig:dpwsa}, the two solid lines, from bottom up, are
inverted solutions for SCDM1 and SCDM2 respectively.  The dotted lines
near them are the inputted 3-D power spectrum for those two models
respectively. For SCDM models, the inverted solutions are still
reasonably close to the original solution, which can be seen clearly
from the figures. It shows that our inversion procedure is robust, and
only mildly dependent on the shape of the power spectrum or non-linear
evolution.

\begin{figure}
\vskip 3.3 truein
\includegraphics{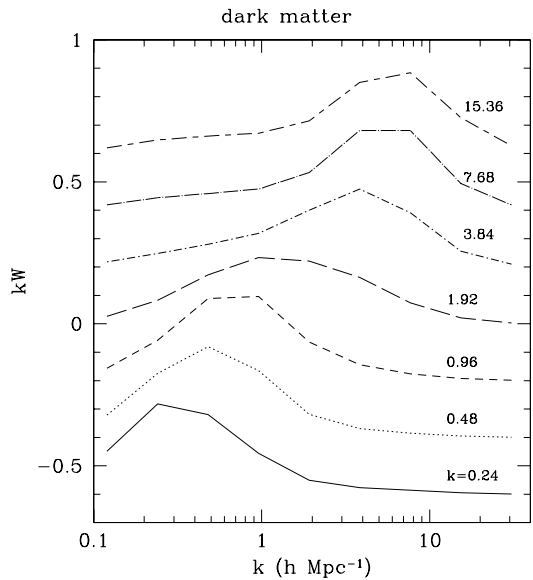}
%\plotone{dwindowa.eps}
\caption{
The normalized window functions of dark matter for method A.
>From bottom to top are for the $k$ bins in figure \protect\ref{fig:dcova}.
For clarity, the lines have been displaced 
down 0.6, 0.4, 0.2, 0, -0.2, -0.4 and -0.6 units respectively.
}
\label{fig:dwindowa}
\end{figure}

As described above, the minimum variance power spectrum estimation
from Method A results in significant smoothing of the input power.
This is described by a window function, which is shown in Figure
\ref{fig:dwindowa}.  For our linear procedures, each bandpower
estimator depends not only on the power in its own band, but due to
geometry also on aliased power from other bands.  The response of the
estimator is just the window function.  We see the increased breadth
of each window compared to the two dimensional window, and there
really are only a smaller number of independent points.  Due to the
low signal-to-noise of our data, we used wide logarithmic bins, and
the window functions are also averaged over the same bin sizes.

\begin{figure}
\vskip 3.3 truein
\includegraphics{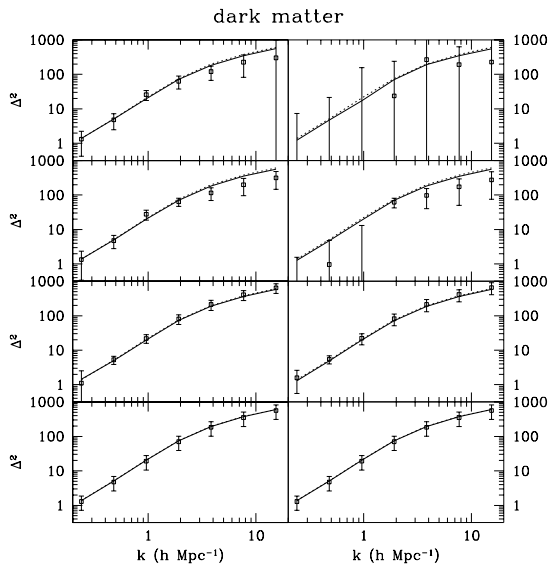}
%\plotone{dpwsbc.eps}
\caption{3-D power spectrum of dark matter at redshift zero.
The left panels are for method B, and the right panels are for method C.
%The meaning of lines and signs are the same as in Fig. \ref{fig:dpwsa}. 
The dotted lines are the input power spectrum, linearly evolved from
the \citet{2002astro.ph..7664S} power spectrum at $z_m$.
The solid
lines are the deprojected 3-D power spectrum from the non-linearly projected
2-D power spectrum.  The boxes are the power spectrum deprojected from
the measured angular power spectrum.
The panels from bottom to top include 1, 3, 5, 7 SVD values in their
reconstruction, respectively. 
}\label{fig:dpwsbc}
\end{figure}

\begin{figure}
\vskip 3.3 truein
\includegraphics{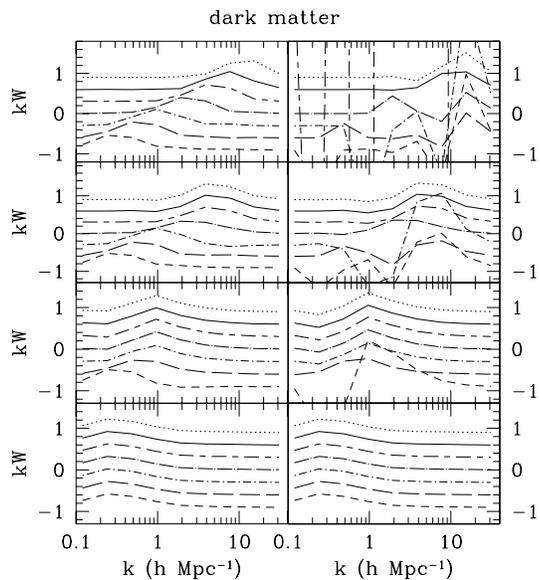}
%\plotone{dwindowbc.eps}
\caption{
The normalized window function of dark matter for different SVD cutoffs.  
The panels
are positioned as in Figure \ref{fig:dpwsbc}, displaced
down 0.9, 0.6, 0.3, 0, -0.3, -0.6, -0.9 units respectively.
At the lowest panel, only one SVD is used, so the solution only has
one degree of freedom and all reconstructed values are linearly degenerate.
}
\label{fig:dwindowbc}
\end{figure}

The goal of methods B and C are to reduce the width of these windows
by deconvolving the solution by the window.  If all steps were
non-singular, method C would result in $\delta$-function windows.  For
the SVD procedure, one chooses the number of eigenvalues (also called
singular values) to include.  The result will depend on the number
of modes included.  In Figure \ref{fig:dpwsbc} we show the recovered
power spectrum as we increase the number of modes from bottom to top.
The corresponding window functions are shown in Figure
\ref{fig:dwindowbc}.  For method C (right panel, Figure
\ref{fig:dpwsbc}), we see that increasing the number 
of modes results in increasing errors, as one expects from
deconvolutions.  For method B (left panel, same figure), the results
are quite robust, and no cutoff is actually needed.  The window
functions in Figure \ref{fig:dwindowbc} show the effect of the SVD
cutoff graphically. The bottom panel shows that if only one mode is used, all 
solutions are degenerate. As one increases the number of modes used, the
windows shift apart.  Method B remains stable, and results in windows
that are narrower than Method A (shown in Figure \ref{fig:dwindowa}).
For method C, the windows become ill-behaved, and one does not obtain
a good window structure for any number of modes.

\begin{figure}
\vskip 3.3 truein
\includegraphics{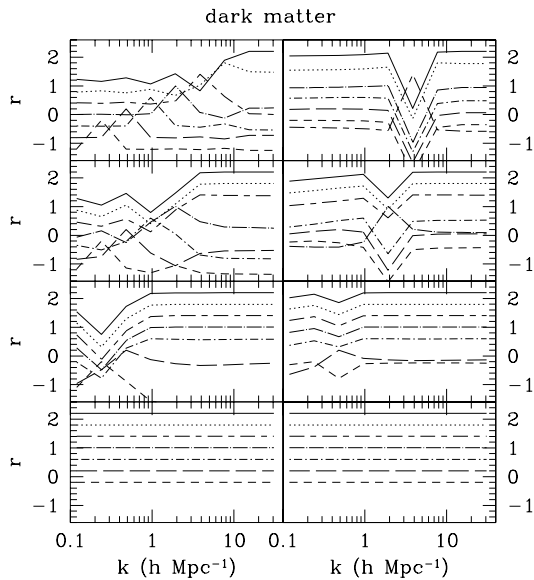}
%\plotone{dcovbc.eps}
\caption{cross-correlation coefficient for the hierarchy of solutions. The 
left panels are for method B, and the right panels are for method C.
The panels from bottom to top include 1, 3, 5, 7 SVD values in their
reconstruction, respectively. 
The successive lines have been displaced
down 1.2, 0.8, 0.4, 0, -0.4, -0.8, -1.2 units respectively.
}
\label{fig:dcovbc}
\end{figure}

The covariance of bins shows the linear structure of the solutions.
As before, we can define the cross-correlation coefficients between
bins.  The cross-correlation coefficients of Method A are shown in
figure \ref{fig:dcova}. For Method B and C, if only one singular value
is used, all modes are linearly dependent, so the cross- correlation
coefficient is unity as can be seen in figure \ref{fig:dcovbc}.
Method B decorrelates the bins, making the statistical interpretation
of the results particularly simple.  The very high correlation between
points in Method C shows that one never really recovers more than 2
independent modes regardless what cutoff one chooses.  The full
deconvolution does not lead to meaningful results.

The top left panel in Figure \ref{fig:dpwsbc} is most readily compared
to the literature.  It is deconvolved but still stable. We see the
three bins centered at $k=0.96$ which have good signal to noise.  The
covariance matrix in Figure \ref{fig:dcovbc} tells us that the three
points are statistically uncorrelated.  The top left panel in the
window function Figure \ref{fig:dwindowbc} has a width of about two
bins.  While one might wonder how the bins can overlap spatially and
still be uncorrelated, this is not a contradiction: the 2D projected
power spectrum sums over many independent 3D modes.  In the
deprojection, each bin can still depend on different modes of the same
absolute wave number.  This gives as a series of statistically independent
estimators of powers which probe the same physical length scales.

\citet{2002PhRvD..66j3508T} plotted a 3D estimate for the linear dark
matter power as probed by RCS.  In the Hamilton paradigm
\citep{1991ApJ...374L...1H}, the non-linear structure on a given scale
comes from the gravitational collapse of a larger scale.  In an
isotropic collapse, one expects the non-linear wavenumber to be given
by the cube root of the density times the linear wave number, $k_{\rm
NL}=(1+\Delta^2(k_{\rm NL}))^{1/3}k$.  Our three best points at
$k_{\rm NL}=0.48,0.96,1.92 h$Mpc$^{-1}$ map to
$k=0.29,0.37,0.53h$Mpc$^{-1}$, which is comparable the converted
linear length scales measured by the RCS data.  We then used the
\citet{1996MNRAS.280L..19P} prescription to map the non-linear power
to a linear power.  The mapping was done relative to the fiducial LCDM model.

\begin{figure*}
%\vskip 3.3 truein
%\special{psfile=qaz.eps hscale=40 vscale=40 angle=0 hoffset= -18 voffset=-50}
\psfig{figure=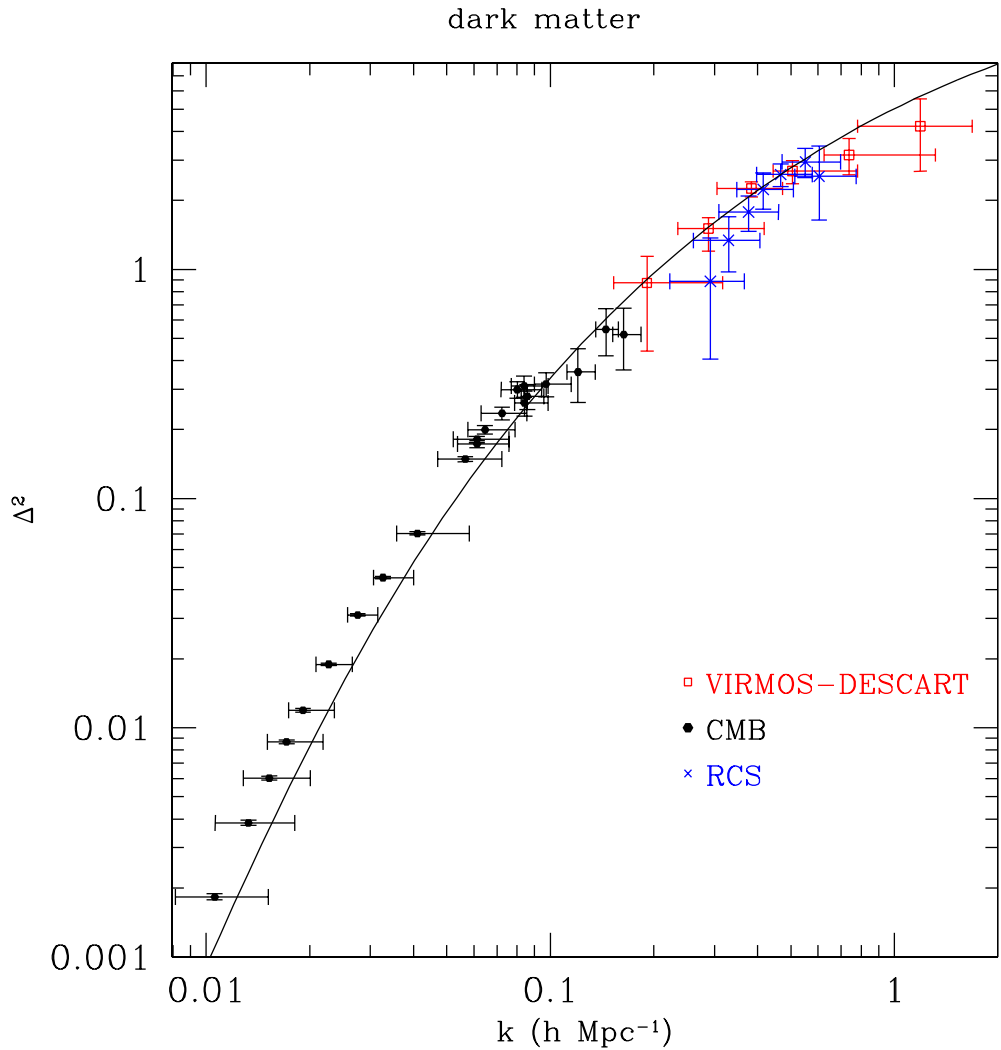,width=15cm,angle=0.}
\caption{
Linearized CMB and RCS comparison power spectrum from
\citet{2002PhRvD..66j3508T}.  The open boxes are the VIRMOS-DESCART
power.  They are mutually uncorrelated.
The solid boxes are a combination of CMB powers.  The crosses
are RCS points.  The lensing data 
sets are linearly decompressed and evolved to $z=0$ (see text for
details).  The solid line is the linear LCDM model.
}
\label{fig:tegmark}
\end{figure*}
The resulting combined CMB and weak lensing data is shown in Figure
\ref{fig:tegmark}.  Our horizontal error bars are derived from the
half width of the window function (top left panel of Figure
\ref{fig:dwindowbc}).  The error bars have been rescaled to 
the \citet{2002PhRvD..66j3508T} convention of 20\% to 80\% using
a Gaussian model.  The CMB data, courtesy of Max Tegmark, includes
all the experiments compiled in \citet{2002PhRvD..66j3508T} as well as the
recent WMAP data (Tegmark, private communication).
We see good agreement between two completely
different lensing data sets (RCS and VIRMOS-DESCART), as well as a
good fit to the standard cosmological model.  The residual differences
could well arise from the subtleties in modelling the PD96
prescription, since we see a better fit to the same model using the
\citet{2002astro.ph..7664S} algorithm in the non-linear power shown in
the top left panel of Figure \ref{fig:dpwsbc}.  Unfortunately the
newer, more accurate non-linear formulae are difficult to invert from
non-linear to linear power heuristically. 

The CMB data can be predicted from first principles to exquisite
accuracy.  The weak lensing is similarly predictable from first
principles, and only limited by the accuracy of simulations.  These
are currently not a limiting step, but do need to improve to match
newer lensing data sets.  Both CMB and weak lensing data sets are
observationally challenging to obtain, but theoretically very clean to
interpret and are unlikely to contain astronomical uncertainties.  We
expect ongoing surveys such as the CFHT legacy survey
(http://www.cfht.hawaii.edu/Science/CFHLS/) to bring the
measurement of non-linear power to a precision era, for which we can
then perform precision cosmology without invoking complex poorly
understood radiative phenomena.

\begin{figure*}
\centerline{
\psfig{figure=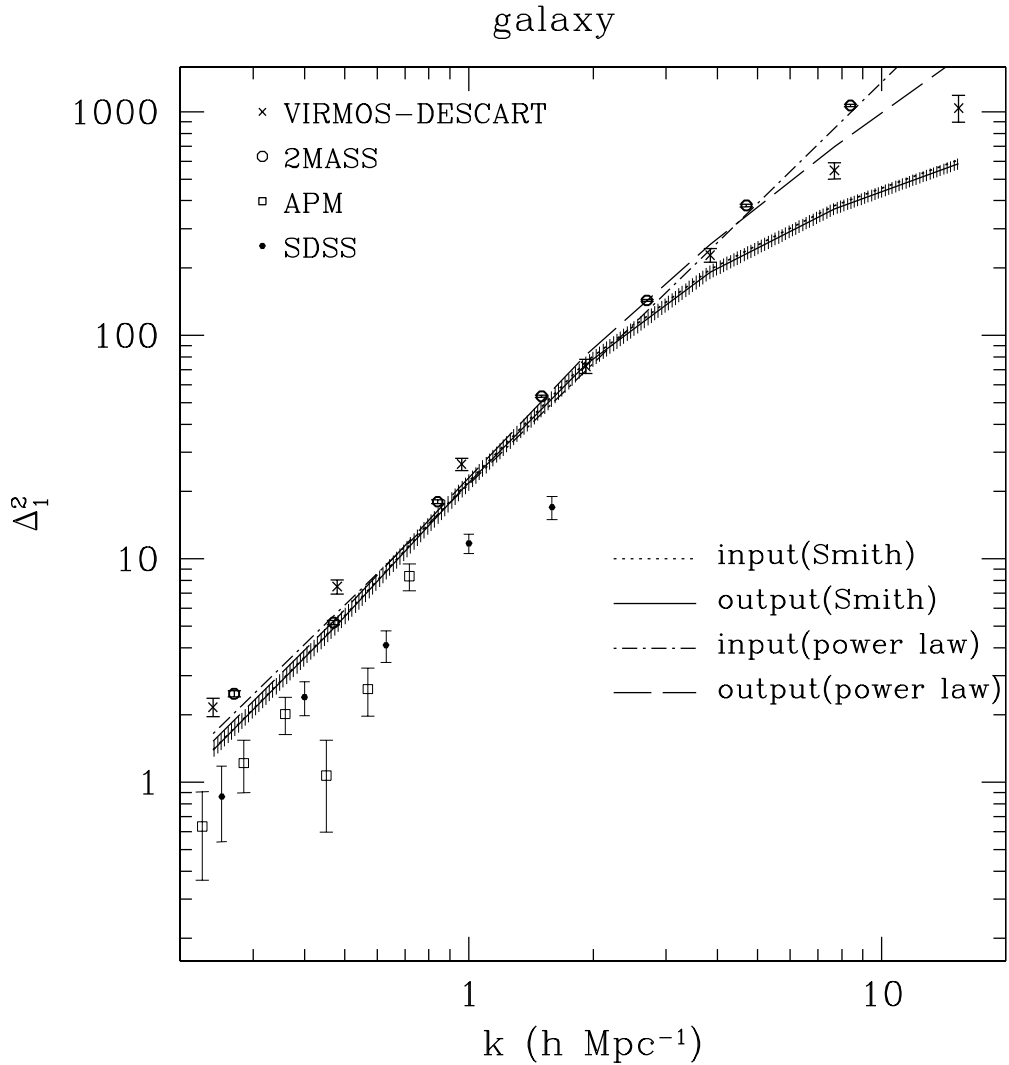,width=15cm,angle=0.}
}
\caption{
The 3-D power spectrum of galaxies at redshift zero, for the minimum
variance method A.
The dotted line is the input power spectrum, linearly scaled from the
 \citet{2002astro.ph..7664S} power spectrum at $z_m=0.36$.
The solid line is the deprojected 3-D power spectrum from the nonlinear
projected 2-D power spectrum.
The crosses are the power spectrum deprojected from
the measured angular power spectrum.
The shaded region is covering the deprojected 3-D power spectrum 
when the input redshift $z_s$ varies over the bootstrap uncertainty
from 0.325 to 0.355.  
Boxes are the power spectrum from APM \citep{2001ApJ...546....2E},
small circles from 2MASS \citep{2003astro.ph..4005M},
and large solid circles
from SDSS \citep{2002ApJ...572..140D}.
The dot-dashed line is the parametrized power spectrum 
$(k/(0.18 h{\rm Mpc}^{-1}))^{1.8}$. 
The dashed line is the projection/deprojection of this power law power
spectrum. 
}
\label{fig:gpwsa}
\end{figure*}

\subsection{galaxy and cross power spectra}
\begin{figure}
\vskip 3.3 truein
\includegraphics{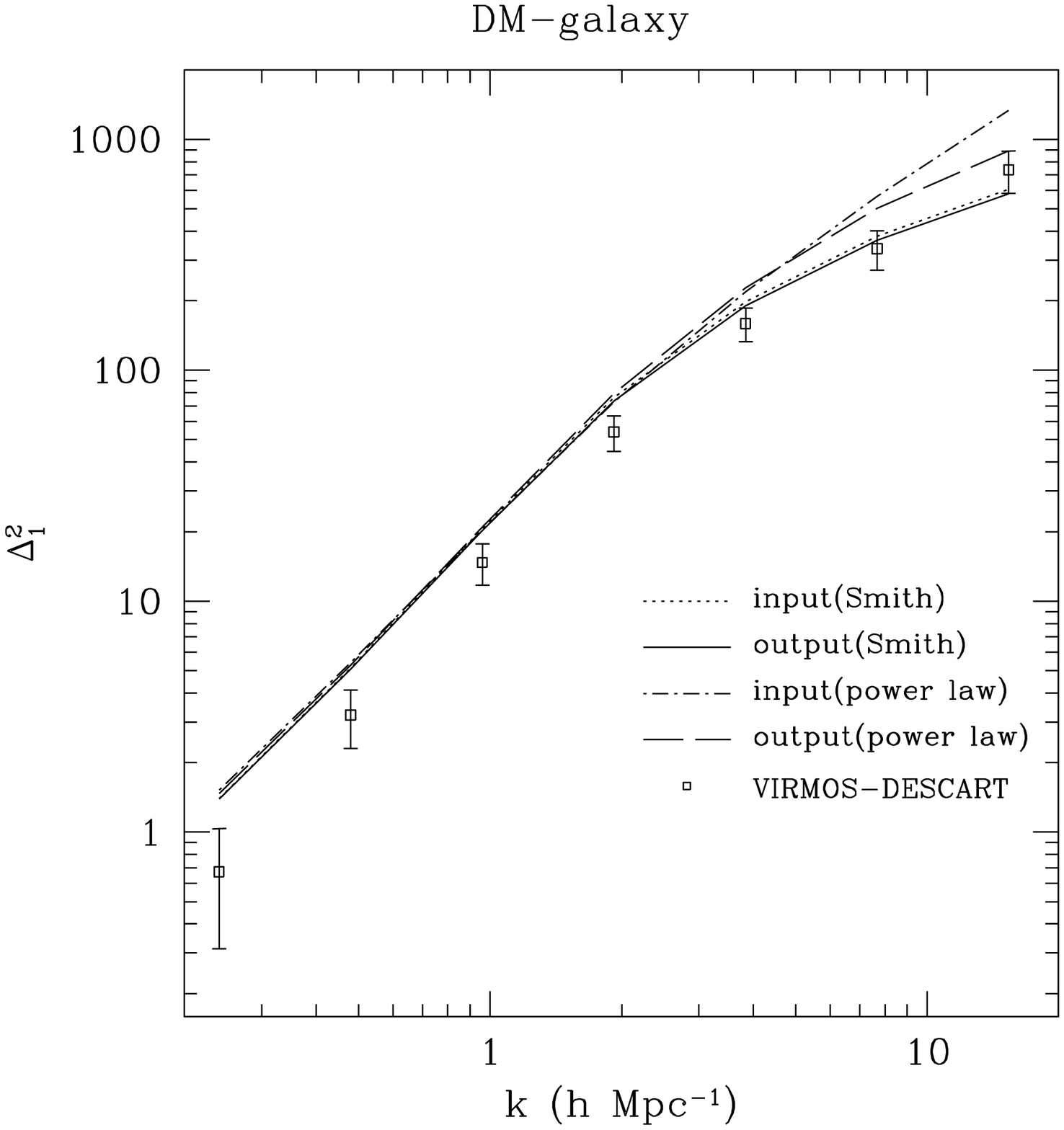}
%\plotone{dgpwsa.eps}
\caption{The 3-D power spectrum of the cross-correlation of galaxy and dark 
matter at redshift zero, for method A.
The box is the power spectrum deprojected from the angular power
spectrum. 
The dotted line is the original power spectrum, linearly evolved from the
\citet{2002astro.ph..7664S}
power spectrum at $z_m=0.36$.
The solid line is the deprojected 3-D power spectrum from the non-linearly 
projected 2-D power spectrum.
The dot-dashed is the cross power for non-stochastic galaxies where
the bias is adjusted to give a power law correlation, 
and the dashed line is the corresponding deprojection.
}
\label{fig:dgpwsa}
\end{figure}

\begin{figure}
\vskip 3.3 truein
\includegraphics{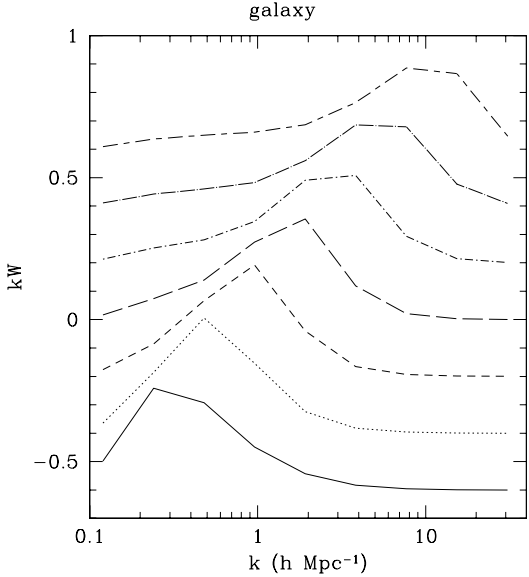}
%\plotone{gwindowa.eps}
\caption{
The window function of galaxies, for method A.
>From bottom to top are for $k=0.24 $,
$k=0.48 $,
$k=0.96 $, $k=1.92 $,
$k=3.84 $,
$k=7.68 $, $k=15.36 (h \rm Mpc^{-1})$, which have been
moved down 0.6, 0.4, 0.2, 0, -0.2, -0.4, -0.6 units respectively.
}
\label{fig:gwindowa}
\end{figure}

\begin{figure}
\vskip 3.3 truein
\includegraphics{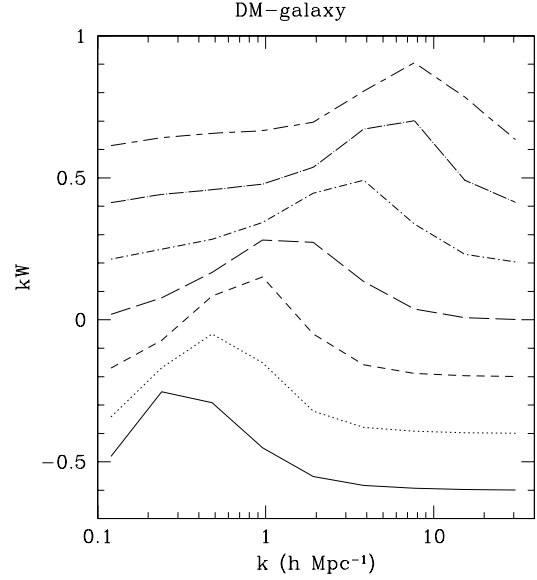}
%\plotone{dgwindowa.eps}
\caption{
Window function of the cross-correlation of galaxy and dark matter,
for method A. The meaning of the lines and signs are the same as
Fig. \ref{fig:gwindowa}.
}
\label{fig:dgwindowa}
\end{figure}

The galaxy and cross power spectra are obtained completely
analogously.  The results for Method A are shown in Figures
\ref{fig:gpwsa} and \ref{fig:dgpwsa}, and corresponding window
functions are Figures \ref{fig:gwindowa} and \ref{fig:dgwindowa}. We
compare our results to that obtained by inverting angular power
spectra from 2MASS \citep{2003astro.ph..4005M}, APM
\citep{2001ApJ...546....2E} and SDSS \citep{2002ApJ...572..140D}.  To
compare with APM, we used table 4 of \citet{2001ApJ...546....2E} and
used their $(C_P^{-1})^{1/2}$ as error bars.  In order to compare with
our result, the power of APM has been linearly evolved from $z=0.11$
to zero in figure \ref{fig:gpwsa}, which is analogous to our analysis.
For SDSS, we use the data in table 2 of \citet{2002ApJ...572..140D}.
We used the error bar not including redshift errors, and linearly
evolved every bin to redshift zero.  The median weight redshifts were
taken from table 1 of \citet{2002ApJ...572..140D} given by the SDSS
photo-z's.  We combined the results of the 4 magnitudes bins weighted
by the inverse variances.

The VIRMOS-DESCART galaxies clearly show more power than the SDSS or
APM data set. For reference, Figure \ref{fig:gpwsa} also shows the
dark matter power (solid line) and the best fit $n=1.8$ power law to
the angular power spectrum (dot-dashed line).  In the inversion, we
used the normalization coefficients ${\bf N}$ of the dark matter,
which can introduce a bias for a power spectrum of different slope.
The result of projecting and deprojecting the power law is shown as
the long dashed curve, which is in general agreement with the input
power. 
 
To put the result in perspective, the best fit correlation length
(described in section \ref{sec:2d}) is $r_0=4.79\pm 0.22 (h^{-1}\rm
Mpc)$ at median redshift $z_m=0.36$ for fixed power law index $n=1.8$.
This is similar to correlation lengths found in CNOC2
\citep{2001ApJ...560...72S}, where the correlation length varied from
$r_0 \sim 3.4-5.5 h^{-1}$ Mpc depending on galaxy population and
redshift.  The different power in APM and SDSS might just be a
reflection of different galaxy types.  The 2MASS galaxies, which are
also infrared selected like the VIRMOS-DESCART, are in better
agreement.  We also note that overestimating galaxy distances
overestimates the inferred power, and the nominal bootstrap
uncertainty is reflected in the hashed region around the solid line in
Figure \ref{fig:gpwsa}. If we estimate the SDSS correlation length to
be 0.3$h$Mpc$^{-3}$ linearly evolved at $z=0$, we infer a correlation
length $r_0=3 h^{-1}$ Mpc, which is at the low range for the CNOC2
sample.  We should note that the overall redshift calibration could be
off by more than the bootstrap error.  The redshift distribution of
galaxies is also broad, and we assumed a linear evolution model for
the clustering, which might not be what the galaxies actually do.  In
principle this evolution can be measured from the data itself by
modelling the galaxies at difference magnitude cutoffs, which is the
subject of a future paper. In any case, this comparison 
puts forwards the central role of redshift information for a correct 
cosmological interpretation of the data

The cross power spectrum from LCDM (table \ref{tab:models}) is shown
in figure \ref{fig:dgpwsa} as the dot-dashed line.  The dashed line
near it is the deprojected power spectrum from the non-linearly
projection of the new cross power. The two lines are still close to
each other, which means the inversion process for the cross power spectrum is
also robust. % The inverted results using Method B and C are shown in

\begin{figure}
\vskip 3.3 truein
\includegraphics{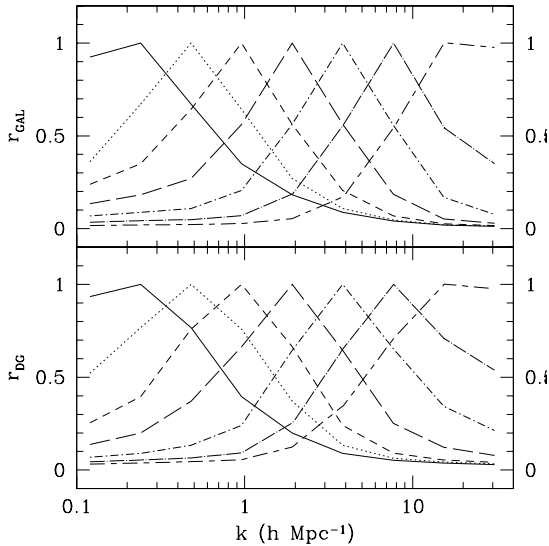}
%\plotone{cova.eps}
\caption{cross-correlation coefficient using method A for the galaxy
and the galaxy-dark matter cross correlation.
The top panel is for galaxy, and the bottom panel is for galaxy-dark matter 
cross correlation.
The meaning of lines are the same as Fig. \ref{fig:dcova}. 
}
\label{fig:cova}
\end{figure}

Just as for dark matter, each of the bins has correlations.  Qualitatively,
they are similarly behaved to that of the dark matter.  We only show the
cross-correlation coefficient for method A in figure \ref{fig:cova}.
%For the galaxy power spectrum, the signal-to-noise is significantly
%larger than for dark matter.  

%As one can see from Figure \ref{fig:gcovbc}, 
%\begin{figure}
%\plotone{gcovbc.eps}
%\caption{cross-correlation coefficient using methods B and C for the galaxy
%power spectrum.  Unlike the dark matter case, the full inversion leads
%to meaningful reconstructed modes due to the larger signal-to-noise.
%}
%\label{fig:gcovbc}
%\end{figure}
%with improved signal-to-noise, deconvolution still results in
%meaningful power spectrum estimates, albeit with significant
%anti-correlations. 

\begin{figure}
\vskip 3.3 truein
\includegraphics{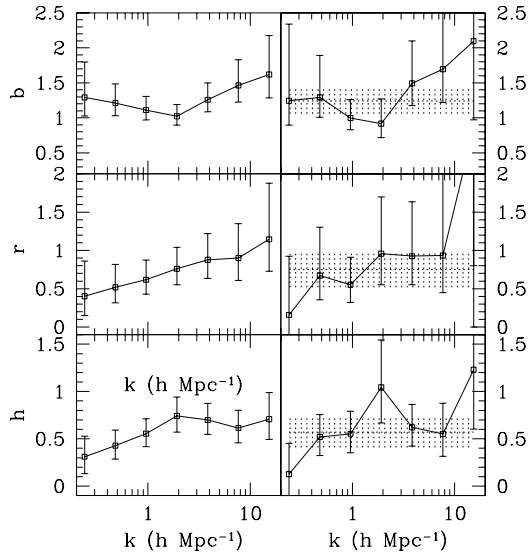}
%\plotone{brh.eps}
\caption{
The top panels are bias of galaxies versus dark matter, for method A and B 
from left to right.
The middle panels are cross-correlation coefficients of galaxy and dark matter,
for method A on the left, and method B on the right.
The bottom panels are the ``galaxy halo parameter''.
The dotted lines in the right panels are best-fit values and the shaded
region show the formal uncertainty:
$b=1.24\pm0.18$, $r=0.75\pm0.23$, $h=0.57\pm0.16$.  In our linear
evolution model, the results are redshift independent.
}
\label{fig:brh}
\end{figure}

With the full set of deprojected three dimensional power spectra of
galaxies  $\Delta^2_{\rm gal}$, dark matter $\Delta^2_{\rm dm}$,
and their cross-correlation  $\Delta^2_{\rm cross}$, we can
directly measure the derived quantities ``bias'' $b$ and
``stochasticity'' $r$ \citep{1998ApJ...504..601P}.
The bias is:
\begin{equation}
b=\sqrt{\frac{\Delta^2_{\rm gal}}{\Delta^2_{\rm dm}}}
\label{eqn:bias}
\end{equation}
which is shown in Fig.\ref{fig:brh}.  The upper error bars are obtained
by using the $1-\sigma$ upper value for $\Delta^2_{\rm gal}$ divided by
the $1-\sigma$ lower value of $\Delta^2_{\rm dm}$, and
analogously for the lower error bar. From the definition, bias is dependent on the cosmology. 
The galaxy-dark matter cross-correlation coefficient is also directly
measurable, 
\begin{equation}
r=\frac{\Delta^2_{\rm cross}}{\sqrt{{\Delta^2_{\rm dm}}{\Delta^2_{\rm
	gal}}}}.
\label{eqn:r}
\end{equation}
The dark matter power has the largest error bar.  One can take the
ratio of cross and galaxy power, which has smaller errors.  We call
this the ``galaxy halo parameter'' $h$, defined as
\begin{equation}
h\equiv\frac{r}{b}=\frac{\Delta^2_{\rm cross}}{\Delta^2_{\rm gal}}.
\label{eqn:h}
\end{equation}
It is shown in the bottom panel of Fig. \ref{fig:brh}. The error bars
are drawn using the procedure described above. This halo parameter is 
also dependent on cosmology, but the galaxy-dark matter 
cross-correlation coefficient is independent. 

We fitted a constant average value for $b$ from the minimum in
$\chi^2$ for Method B: 
\begin{equation}
\chi^2=\frac{1}{N}\sum_{i=1}^N{\frac{\left[\Delta^2_{\rm gal}(k_i)
-b^2 \Delta^2_{\rm dm}(k_i)\right]^2}{\sigma^2_{\rm gal}(k_i)
+b^2\sigma^2_{\rm dm}(k_i)}}
\end{equation}
sampled at six wavenumbers $k_i=$0.24, 0.48, 0.96, 1.92, 3.84,
7.68$h$Mpc$^{-1}$. 
Since the covariances are neglible, we neglected them.
The variance of $b$ is taken from
$\Delta\chi^2$,
\beq
\Delta\sigma_{\chi^2}
=\frac{\sqrt{2}}{\sqrt{N}}=\frac{1}{2}\frac{d^2\chi^2}{db^2}(\delta b)^2.
\eeq
We apply the same procedure to solve for $r$ and $h$.  Formally, we find
$b=1.24\pm0.18$, $r=0.75\pm0.23$, and $h=r/b=0.57\pm0.16$ with $\chi^2=1.20,
0.51,0.86$ per d.o.f. for six degrees of freedom, consistent with
the expected standard deviation of 0.6.
Just as the bias measurements are subject to systematic redshift
calibration uncertainties, the cross-correlation results could also
depend on such issues.  Future surveys with photometric redshifts
signficantly reduce this problem.

The resulting bias $b$ is larger than that found from comparing
VIRMOS-DESCART to RCS \citep{2002ApJ...577..604H}.  That study found
$b=0.71^{+0.06}_{-0.04}$ and $r=0.57^{+0.08}_{-0.07}$.  The cross
correlation coefficient $r$ is consistent.  The apparent discrepancy
in the bias $b$ should not be overinterpreted.  This earlier
comparison was in 2-dimensional projection.  Furthermore, RCS galaxies
are red R selected, which can be a different population.  Within the
SDSS galaxies on our scales of $k\sim 1 h$Mpc$^{-1}$ the power for
galaxies varies by a factor of 4 between the 18-19 and the 21-22
magnitude bin.  Similarly, the 2MASS power \citep{2003astro.ph..4005M}
is closer to our derived value than the R selected SDSS galaxies.  One
clearly needs to exercise care when converting the parameter fits from
one sample of galaxies to another one, as galaxies selected in
different colours will have quite different clustering properties.

Our bias, cross-correlation and halo parameters were all estimated in
3-D space.  One could have also attempted a parametric estimation on
the 2-D projected power.  For an optimal inversion process if one takes
the covariances between scales into account, the results will be the
same.  The single biggest source of larger error is that we added the
$B$ mode power to the error budget.  Our model of sample variance in
the optimal 2-D power will also give a larger sample variance error.
The reason for that is the under-estimate of error on the 2-pt
correlation function.  The previous sample variance errors were
estimated using an effective contiguous area.  The masks and source
clustering will increase the sample variance, since the same area is
now non-uniformly sampled.  In Figure \ref{fig:clvirmos}, one sees
that at $l$ up to 1000, sample variance makes up half the error budget
(the dashed line is the noise, which exceeds the signal at $l\sim
1000$).  At the end of the day we do have about twice the
error bar, coming from this combination of factors.

We had checked the effects of redshift binning, and at 15 there was
about a percent change compared to an infinite number of slices.  The
redshift evolution is parametrized, so the finer the redshift bins are
the more accurate it gets.  In the linear evolution model, the
redshifts scale the same for the galaxies and dark matter, so that
cancels exactly and is redshift invariant.

One expects galaxies and dark matter to be well correlated on linear
scales when $\Delta^2<1$ for both galaxies and dark matter, which is
not well probed by the angular scales of the current data.  Newer larger
surveys should significantly improve on the angular scale coverage.
When the two fields are well correlated, there is no sample variance
in the measurements of $r$ and $b$, which is reflected in the full
joint estimation Fisher matrix.  Using the current data, however, the
errors in large scales are dominated by a $B$ mode, which is not
easily modelled.

\section{Comparison with Theory}

In this section we will discuss the results of the study in the
context of theories and other surveys.  Measuring the relation between
the distribution of light and that of dark matter has significant
cosmological consequences, as discussed in the introduction.  In the
other direction, the theory of galaxy formation requires observational
constraints to be tested.  While physical cosmology originated
thirty years ago in the paradigm that stars account for all the mass
in the universe, today's picture is very different.  The universe
appears dominated by very mysterious dark energy accounting for about
70\% of the energy density of the universe.  The second most important
energetic contribution is dark matter, accounting for another 27\%.
Ordinary baryonic matter accounts for another 3\%.  The visible stars
account for less than 0.3\%.  Optical power spectra of galaxies
measure the distribution of this 0.3\% of matter, which may or may not
be a good tracer of the hundredfold more abundant dark matter.  The
challenge to galaxy formation models is to understand the distribution
and kinematics of that small fraction of visible stars.

Different galaxies are composed of different stellar populations.
Galaxies of different types cluster differently.  Empirically it is
known that red (early) type galaxies cluster more strongly than blue
(late) type galaxies.  The goal of the theory of galaxy formation is
to quantify the distribution of visible galaxies, i.e. the
distribution of visible light.  This distribution is quantified by
various statistical properties.  At the two point level, theories of
galaxy formation can be tested by predicting the two point statistics
measured in this paper: auto and cross correlations.  This correlation
is a function of colour, morphology and redshift.

Semi-analytic studies generically predict \citep{2001MNRAS.320..289S}
earlier type (red elliptical) galaxies to be more strongly biased than
late type (blue spiral) galaxies.  Qualitatively, one might expect the
VIRMOS-DESCART $I$ band selected galaxies to be systematically
redder than SDSS or RCS galaxies, and therefore more clustered
(i.e. more biased).  Here we should keep in mind that the restframe
colours at our median weighted redshift $z_m\sim 0.36$ are
significantly bluer, so RCS/SDSS $R$ bands are closer to restframe
$V$, while VIRMOS-DESCART $I$ shifts into the rest frame $R$.
Comparison with APM or 2MASS is furthermore complicated by the
significant difference in redshift distributions: these latter two
surveys are much shallower with median redshifts of 0.11 and 0.07
respectively.  In our comparison plot shown in Figure \ref{fig:gpwsa},
the different redshifts were scaled using a linear evolution model.
Linear evolution assumes that clustering increases due to
gravitationally induced motions.  In biasing models, the clustering is
enhanced by creating the objects in a more clustered fashion, such
that the gravitational motions have a smaller fractional effect.
A biased population is expected to evolve more slowly.
The 2MASS galaxies actually have significantly more power than our
VIRMOS-DESCART sample.  Should the picture hold that the redder surveys
select for earlier type galaxies, \citet{2001MNRAS.320..289S} predict
an increasing bias as one goes from blue to red, which is from APM to
RCS/SDSS to VIRMOS-DESCART to 2MASS.  The stochasticity as
parametrized by the cross correlation coefficient $r$ was predicted to
be less population dependent, which is what we observe.

These qualitative statements are not easy to quantify.  One would need
to have a common measure of galaxy morphological classification into
early and late types, and correct for evolutionary effects.
Empirically, a careful study of the CNOC2 survey
\citep{2001ApJ...560...72S} showed the strong dependence of the
clustering amplitude on the galaxy types.  The relative evolution of
each population was rapid, and the early types were much more
clustered than the late types. 

\citet{2000astro.ph.11458P} did a morphological breakdown of the
2MASS population, and also a comparison to several other surveys.
According to their estimates, the bright 2MASS sample consists of a
mixture of about half early and half late type galaxies, while APM and
other blue selected surveys have 20\% early 80\% late type galaxies.
This is likely the origin for the larger power in the 2MASS survey,
especially on small nonlinear scales.  \citet{1998ApJ...499..112B}
measured the CFRS galaxy morphologies with HST, and also find about equal
early and late type galaxies in our magnitude range.  The CFRS galaxies
are selected by similar colours as VIRMOS-DESCART, so we expect
2MASS and VIRMOS-DESCART to yield similar results.  The difference
in redshift distribution opens up some leeway, but our results are in
general consistent.

The observable statistics do not end at the two point function.  The
full three point function and windowed skewness for the dark matter
has been measured to better than 10\% accuracy
\citep{2003astro.ph..2031P,2002A&A...389L..28B}.  The cross skewness
to luminous matter has 4 moments \citep{1998ApJ...504..601P}, which
are all directly measurable and provide additional constraints on
galaxy formation models.

We note at this point that our model for power spectrum inversion was
designed with dark matter evolution in mind, which is physically well
understood from first principles.  We naively applied this model to
the galaxy and cross-correlation power using the same assumptions,
that light traces mass and that the stochasticity is small.  Our
results obtained under these assumptions show that the assumptions are
only partially true: the optical galaxies are biased, and there is
evidence for stochasticity.  The galaxy halo parameter $h$ was
inconsistent with unity, ruling out a mass traces light model.  When
light does not trace mass, as we have found, our linear evolution
model used to deproject the galaxies is not a unique interpretation of
the galaxy power.  One could have many different plausible mechanisms
which can lead to the same observed data set, but with quite different
underlying properties.  Even within a magnitude range, one is
measuring a mixture of nearby intrinsically faint galaxies and distant
intrinsically bright galaxies.  It is likely that these two
populations have different power, and do not evolve by any simple
parametrized model.  In the future, photometric redshifts will allow
separation of several of these effects, and give a systematic
parametrized hierarchical measure for galaxy formation.

\section{The Galaxy-Dark Matter Connection}

Mathematically, all two point statistics encode the same information.
When we observe the distribution of dark matter and galaxies, all two
point information is described by two power spectra and one cross
spectrum.  One can also construct derivatives of these quantities, and
non-linear combinations, for example the bias and cross correlation
coefficient shown in the previous sections.

Historically, the paradigm to understand dark matter was not on this equal
footing, but rather centered on visible galaxies.  One could measure the
luminosity and number density of galaxies.  A popular strategy was to
attempt to measure the mass concentration associated with the light.
The mass concentration is known to have a larger spatial extent than the
light, which was parametrized as the radial halo mass distribution
which we call the ``halo profile''.
If one could measure all the mass associated to halos, one could in
multiply the number density of galaxies by the mass of the halos to measure
the mass in the universe.  Of course there could also be mass that is
not associated with visible galaxies, so that would still only represent
a lower limit to the mass density of the universe.
We can connect these two viewpoints, which are different interpretations
of the same numbers.

Several approaches exist to measure the halo mass.  Perhaps the most
direct is galaxy-galaxy lensing.  One defines a halo profile $H(r)$,
and considers each galaxy to have a halo, such that the dark matter
distribution is the convolution of the position of galaxies described
by a density distribution $\rho_{\rm gal}$, which may be a sum of
$\delta$-functions.  One further assumes that all dark matter is
associated with such halos.  Galaxies of different morphologies,
luminosity or colour may have different halos, and one can do a full
segregated measurement.  The formalism remains the same, so we only
consider one universal galaxy class.  One then tries to fit for this
universal halo profile.  Apart from noise weights, all analysis
proceed as follows.
The distribution of ``halo'' dark matter is
\beq
\rho_{\rm halo}\equiv
\int \rho_{\rm gal}(x') H(|x-x'|) d^3x'.
\label{eqn:gdm}
\eeq
If one sums the tangential shear to each galaxy, and stacks all such
galaxies, one has cross correlated the dark matter field
(\ref{eqn:gdm}) with the galaxy field,
\beq
\xi_{t}(r)=\frac{\langle\rho_{\rm gal}(x) \rho_{\rm
    DM}(x+r)\rangle}{\bar{\rho}_{\rm gal}\bar{\rho}_{\rm DM}}.
\label{eqn:xigg}
\eeq
In the halo model, one equates $\rho_{\rm halo}= \rho_{\rm DM}$. 
Formally, this yields the cross-correlation of the galaxy position with
the associated halo mass.  This is equivalent to 
Fourier transforming, and multiplying both sides of equation
(\ref{eqn:gdm}) by the galaxy density,
\beq
\Delta^2_{\rm cross}(k)=\Delta^2_{\rm gal}(k)h(k).
\eeq
We have absorbed the normalization coefficients into the dimensionless
halo profile $\bar{n} h(r)\equiv H(r)$ with
$\bar{n}=\bar{\rho}_{\rm DM}/\bar{\rho}_{g}$.
Inverse Fourier transforming and applying equation (\ref{eqn:h}) we find  
\begin{equation}
H(r)\equiv \bar{n} h(r)=4\pi \bar{n}\int \frac{r(k)}{b(k)}
\frac{\sin(k r)}{kr} k^2 dk. 
\label{eqn:halo}
\end{equation}
The halo profile is mathematically equivalent to the Fourier transform
of the 
cross correlation coefficient divided by the bias, and requires only
measurements of the galaxy auto-correlation and the galaxy-dark matter
cross correlation functions.  It is the Fourier transform of our
``galaxy halo parameter'' $h$ defined in equation (\ref{eqn:h}).

Formally, one could derive a mass from the halo profile $H(r)$, and
compare this to masses of halos.  This interpretation is parameter
dependent, since the normalization in (\ref{eqn:halo}) depends on the
mean density of galaxies, which depends on the flux limits of the
survey.  The deeper a survey looks, the more faint galaxies one finds,
so the gravitating mass is divided by more galaxies.  Similarly it is
tricky to extrapolate (\ref{eqn:halo}) to obtain a value of the total
cosmological density.  One only measures the mass correlated with
galaxies, and there could be more mass that does not correlate.  I.e.,
if $r<1$, the halo may underestimates total mass.

Since all two point statistics are equivalent, measurement of
``halo profiles'' always measure the mass-light cross correlation.
Fitting to a universal profile is mathematically equivalent to
equation (\ref{eqn:halo}).  If one knows the mean number density
of galaxies, one could use this relation to infer the mean density
of matter $\bar{\rho}_{\rm DM}=\bar{\rho}_{\rm gal} 4\pi\int h(r)
r^2dr=\bar{\rho}_{\rm gal} h(k=0)$ \citep{2001ApJ...555..572W}.  We see
that the results can under or overestimate the total matter, depending
on the the correlation properties of galaxies and dark matter.  For the
fiducial $\Omega_0=0.27$ model used in our analysis, the VIRMOS-DESCART
halo parameter is $h=0.57\pm 0.16$, implying that about $\Omega_{\rm
halo}=h\Omega_0\sim 0.15$ is in dark matter correlated with $I$ selected
galaxies.  This is consistent with the low inferred values of $\Omega_{\rm
halo}$ in the literature \citep{2001ApJ...555..572W,2002MNRAS.337..774S}.

A popular interpretation of the dark matter distribution has been in
terms of halo models \citep{2002MNRAS.335..311G}.  When describing the
distribution of galaxies, one also needs to specify the cross
correlation of galaxies relative to these halos.  These cross
correlation coefficients must be calibrated to the observed values of
the ``galaxy halo parameter'' $h(r)$.

\section{Conclusions}

We have presented the first full optimal analyses of two dimensional
and deprojected 3-D power spectra of dark matter and galaxies.  We
used weak gravitational lensing from the VIRMOS-DESCART survey, and
their relation to the galaxy distribution in the same data set.  The
survey is sensitive to $0.37<k<2.88 h{\rm Mpc}^{-1}$ and probes the
regime of non-linear clustering. We have compared the results of three
different deprojection procedures, and found that our method B based
on partial deconvolution is simultaneously robust, has narrow window
functions, and mostly uncorrelated error bars.  Full inversion is
unstable, as might be expected, and a cut with SVD results in a highly
tangled window function.  No choice of cutoff leads to a meaningful
three dimensional power spectrum for this full inversion, as measured
by the covariance matrix of the solution.  For the galaxy power
spectrum, the inversion is more stable, but still noisy.

We tested the effects of incorrect model priors.  While the true power
is very non-linear, using a linear evolution model relative to the
median redshift does not introduce a large error.  For the dark matter
power spectrum, the errors are dominated by the B-mode and shot noise.
For the galaxy and cross correlation power spectrum the noise is
significantly smaller, and redshift distribution errors dominate.

The deprojected dark matter distribution is consistent with that expected
from the standard WMAP $\Lambda$ cosmologies with $\sigma_8=0.9$.
The results agree well with deprojected RCS lensing data and the CMB
linearly evolved power spectrum.  The galaxy distribution is similar to
the 2MASS galaxies, but more clustered than that found by APM and SDSS.
This may be due to the different colour band selection.  Using the cross
correlation we confirm earlier results that galaxies and dark matter
are indeed distributed stochastically on non-linear scales.

We have quantified the bias and cross-correlation coefficient 
$b=1.24\pm0.18$, $r=0.75\pm0.23$.  A less noisy combination of the
parameters that can be measured is the ``galaxy halo parameter''
$h=r/b=0.57\pm 0.16$.  These parameters describe the relation between
dark and luminous matter, and are the key uncertainties in the
interpretation of galaxy-galaxy lensing.

All error analyses used Gaussian assumptions.  Most of the regime is
noise dominated, for which Gaussianity is a good approximation.
For the sample variance this potentially underestimates
errors \citep{2001ApJ...554...67H}.  We plan future analyses
to quantify this effect using N-body simulations.  The upcoming
Canada-France-Hawaii-Telescope Legacy Survey weak lensing survey should
significantly improve on all measurements, and probe to larger scales.
At larger scales one will be able to measure bias and stochasticity
without being affected by finite field size sample variance.

\section*{ACKNOWLEDGEMENTS}

We would like to thank Chen Xuelei for helpful discussions and input
in the early stages of the project and Max Tegmark for providing the
data for figure \ref{fig:tegmark}.  We also thank Henk Hoekstra, Francis
Bernardeau and Ari Maller for discussions. The research was supported in
part by NSERC and computational infrastructure from the Canada Foundation
for Innovation.

\bibliography{mybib,penbib}
\bibliographystyle{mn2e}

\appendix

\bsp

\label{lastpage}

\end{document}